\documentclass[12pt]{iopart}

\usepackage[dvips]{graphicx}
\usepackage{epsfig}
\usepackage{color}
\usepackage{amssymb}
\usepackage{amsthm}

\begin{document}

	\title{Cooperative effects in nuclear excitation with coherent x-ray light}
	
	\author{Andr\'e Junker, Adriana P\'alffy and Christoph H Keitel}
	\address{Max-Planck-Institut f\"ur Kernphysik, Saupfercheckweg 1, 69117 Heidelberg, Germany}
	\ead{Palffy@mpi-hd.mpg.de}

	\date{\today}

%%%%%%%%%%%%%%%%%%%%%%%%%%%%%%%%%%%%%%%%%%%%%%%%%%%%%%%%%%%%%%%%%%%%%%%
\begin{abstract}	

The interaction between super-intense coherent x-ray light and nuclei is studied theoretically. One of the main difficulties with driving nuclear transitions arises from the very narrow nuclear excited state widths which limit the coupling between laser and nuclei. In the context of direct laser-nucleus interaction, we consider the nuclear width broadening that occurs when in solid targets, the excitation caused by a single photon is shared by a large number of nuclei, forming a collective excited state. Our results 
show that cooperative effects mostly contribute with a modest increase to the nuclear excited state population except for the case of $^{57}\mathrm{Fe}$ where the enhancement can reach almost two orders of magnitude. Additionally, an update of previous estimates for nuclear excited state population and signal photons  for x-ray lasers interacting with solid-state and ion beam nuclear targets taking into account  the experimental advances of the x-ray coherent light sources is given. 
The presented values are an improvement by orders of magnitude and are encouraging for the future prospects of nuclear quantum optics.

\end{abstract}
	\maketitle
%%%%%%%%%%%%%%%%%%%%%%%%%%%%%%%%%%%%%%%%%%%%%%%%%%%%%%%%%%%%%%%%%%%%%%%

%%%%%%%%%%%%%%%%%%%%%%%%%%
\section{Introduction}
%%%%%%%%%%%%%%%%%%%%%%%%%%

For a long time the direct interaction of photons with nuclei was generally considered too small to be relevant, primarily based on estimates about the magnitude of interaction matrix elements as in Ref. \cite{Matinyan}. Nevertheless, first interesting effects were observed in M\"ossbauer spectroscopy experiments, where despite low excitation rates,  single gamma photons electromagnetically induced transparency (EIT) was observed \cite{Olganticross} and coherent control schemes were suggested \cite{Olgacohcontr}.  Nuclear excitation experiments conducted at synchrotron radiation (SR) facilities evolving from the direction of M\"ossbauer spectroscopy have shown that interesting coherent phenomena such as quantum beats and   photon echoes \cite{MoessbauerSpectroscopy} may occur even without exciting a large part of the nuclear sample  \cite{Synchrotron_Shvydko,Synchrotron_Roehlsberger}.
Furthermore, it was observed that resonant scattering of SR on a nuclear ensemble,  such as identical nuclei in a crystal lattice that allows for recoilless nuclear transitions, may occur via an excited state which is excitonic in nature \cite{HannonTrammell,Kagan,Synchrotron_Roehlsberger}. 
The decay of this collective nuclear excited state then occurs coherently in the forward direction, giving rise to nuclear forward scattering (NFS), and in the case of nuclei in a crystal also at  Bragg angles  \cite{Kagan,HannonTrammell,Smirnov1996}.   The correlation of nuclear excitation amplitudes in the excitonic state furthermore leads to 
a speed up of the nuclear decay,  also called coherent decay, which becomes considerably faster than the spontaneous  de-excitation (characterized by the natural lifetime of a single nucleus). Most recent NFS experiments with SR used this feature to measure the collective Lamb shift in nuclei and to demonstrate  EIT with resonant nuclei in a cavity  \cite{Ralfsci,Ralfnat}.
 The aforementioned cooperative effects, although concerning nuclear excitation, have been however historically more related to solid state physics and their impact on the interaction between lasers and nuclei has so far been disregarded.

With the advent and commissioning of new light sources of  higher power, brightness and temporal and transverse coherence, the driving of nuclear transitions with photons was set on the new course of nuclear quantum optics. The direct interaction between nuclei and coherent radiation from x-ray  free electron lasers (XFEL) was proposed for the study of  phenomena well known from atomic systems such as  coherent trapping,  electromagnetic induced transparency or  optical
measurements of nuclear properties such as transition frequency and dipole moment \cite{NuclearQuantumOptics}. This first pioneering work was followed by further studies of the resonant laser-nucleus interaction involving nuclear electric dipole-forbidden transitions \cite{Adriana_DipoleForbidden,Adriana_Photoexcitation} which had been that far traditionally disregarded in presumed analogy to atomic quantum optics. Nuclear coherent population transfer in a stimulated Raman adiabatic passage reminiscent of atomic quantum optics has been also investigated \cite{NSTIRAP}, as well as the 
direct laser-driven quantum nuclear control in a theoretical framework \cite{Rabitz}. In all these works, the narrow bandwidth of nuclear transitions is a limiting issue.  With values being on the order of $10^{-5}..10^{-10} \, \mathrm{eV}$ for a single nucleus, their size lies tremendously below current XFEL energy bandwidths, suppressing the nuclear interaction with the radiation. Basically, the amount of nuclear excitation is dependent on the number of resonant laser photons, and unlike a typical situation in atomic quantum optics, in the case of nuclei the laser bandwidth is orders of magnitude larger than the resonance.  Control of the nuclear bandwidth would very much help the matter.

In this work, we consider the impact of the cooperative excitation on the efficiency of nuclear quantum optics applications. In solid state targets consisting of M\"ossbauer nuclei, the excitation caused by a single photon may be shared by a large number of scattering centers, forming a collective excited state with a sometimes significantly higher decay width than that of a single nucleus. This phenomenon may be exploited in order to enhance the widths of nuclear transitions, and to the best of our knowledge has so far never been included  in calculations of the laser-nucleus interaction.  We show that for certain cases this enhancement of the nuclear transition width can increase by a factor proportional to the number of nuclei in the sample, and thus with the sample thickness within the laser Rayleigh length. We also study the physical limitations of the previous statement. Furthermore, a comparison between the cases with and without cooperative effects is drawn for the excited state population of the $^{57}_{26} \mathrm{Fe}$ isotope at different sizes of the photon beam focal spot and target thickness.

The use of solid-state targets that allow for cooperative effects for laser-nucleus interaction is restricted to nuclear transitions with energies below the laser photon energy available. At present, this value lies in the range of tens of keV. The first operational XFEL worldwide, the  Linac Coherent Light Source (LCLS) at SLAC \cite{LCLSatSLAC,slac.nature,slac.prl}, provides since 2009 laser beams with photon energy of approx. 10 keV (tunability up to 1.2 {\AA} wavelength was reported \cite{slac.nature}). Beam diagnosis on the second and third harmonics of the primary beam has been performed \cite{ratnersecond}. 
The second operational XFEL worldwide, the SPring-8 Angstrom Compact free electron Laser (SACLA) in Japan, has recently achieved the shortest wavelength of 0.634 {\AA} x-rays (photon energy approx. 19.5 keV) \cite{Sacla,SaclaNatPhot}. The European XFEL, at present still in construction at DESY in Hamburg, Germany, will achieve 24.8~keV photon energy, corresponding to a wavelength of 0.5~{\AA} \cite{EuropeanXFELMachineParameters,EuropeanXFELWavelength}. Apart of the photon energy, for nuclear quantum optics applications, a crucial feature of the laser light pulses is the temporal coherence, i.e, the lack of phase jumps throughout the pulse duration. Here XFELs have the potential to bring a significant improvement compared to SR, besides the fact that they are also considerably brighter.  To this day, even XFELs are not able to provide fully coherent laser pulses, but there are already several proposals how to solve this problem in the near future \cite{sxfel.Feldhaus,sxfel.Saldin,sxfel.geloni,XFELO}. Ideas include providing a single pass XFEL with coherent seeding radiation (Seeded XFEL \cite{sxfel.Feldhaus,sxfel.Saldin,sxfel.geloni}) or designing cavities with the help of diamond mirrors \cite{mirror1,mirror2} to allow for multiple passes of the light through the electron beam (XFEL Oscillator \cite{XFELO}).

With these new developments for temporally coherent XFEL pulses and given the tremendous progress in x-ray beam focusing (recently  reduced to a focal diameter of $7 \, \mathrm{nm}$ \cite{XRayFocus}), our previous estimates for the magnitude of excited state population and signal rates based on experimental parameters dating back to 2007 and 2008 are in need of revision. Here we also give an update to the values presented in Refs.~\cite{Adriana_DipoleForbidden,Adriana_Photoexcitation} for both excitation of nuclei in solid-state targets and of nuclei in ion beams. The latter are useful when nuclear transitions with excitation energies higher than the available XFEL photon energies are required. In this case, moderate target acceleration has been proposed to bridge the gap between  x-ray photon frequencies and nuclear transition energies \cite{NuclearQuantumOptics}. 
 We find that the excited state population values are enhanced in comparison to the older estimates by several orders of magnitude in many cases, especially for solid state targets. For a focus of 7~nm, for instance, the complete nuclear population inversion could be reached. Further experimental developments in this field may clear the way for unprecedented possibilities  involving the direct interaction of radiation fields with nuclei. In particular,  nuclear Rabi-oscillations  raise hope for the coherent control of nuclear excited state population, thus having the potential to open  the entire field of nuclear quantum optics.

This work is organized as follows: in Sec. \ref{sec:TheoreticalDescription} we give a brief introduction to the theoretical framework for laser-nucleus interaction and cooperative effects in nuclei. The optical Bloch equations and the form of the laser-nucleus interaction Hamiltonian for higher multipole transitions are derived. Furthermore, we discuss the effective laser parameters entering the calculation. Sec. \ref{subsec:CollectiveEffects} introduces the collective nuclear excitation for M\"ossbauer  nuclei in a lattice and investigates the origin of line width broadening of excited state transitions due to collective effects. In Sec. \ref{sec:NumericalResults} we present our numerical results for a number of nuclei with transition energies between 1 and 100 keV. We conclude with a short summary and outlook.

%%%%%%%%%%%%%%%%%%%%%%%%%%%%
\section{Theory}\label{sec:TheoreticalDescription}
%%%%%%%%%%%%%%%%%%%%%%%%%%%%
In this section, the density matrix formalism is applied to a two-level nuclear system which is resonantly driven by a super-intense electromagnetic field following the outline presented in Ref.~\cite{Adriana_DipoleForbidden}. We  use a semi-classical description, where the nucleus is considered as a hyperfine-split quantum two-level system and the electromagnetic field is treated classically. The effective laser parameters entering the calculation are introduced.  In Sec. \ref{subsec:CollectiveEffects} we introduce the cooperative effects that appear in solid state targets. The nuclear width enhancement is studied by means of an iterative wave equation for the electric field of the scattered radiation.

\subsection{Density matrix formalism for nuclei} \label{subsec:QuantumDynamics}
%%%%%%%%%%%%%%%%%%%%%%%%%%%%%%%%%%%%%%%%%%%%%%%%%%%%%%%%%%%%%%%%%%%

We consider a nuclear two-level system consisting of the ground state $|g\rangle$ and the excited state $|e\rangle$. In the presence of an intrinsic or external magnetic field the two levels will be split in several ground state magnetic sublevels $|I_g M_g\rangle$ and excited state ones $|I_e M_e\rangle$, where $I$ denotes the nuclear spin quantum number and $M$ its projection on the quantization axis.  All nuclei are assumed to initially populate the ground state sublevels. Starting from time $t=0$ they are irradiated with the intensity $\mathcal{I}(t)$. The dynamics of the density matrix $\hat{\rho}$ is determined by \cite{Scully}
\begin{equation}
	i \hbar \frac{\partial}{\partial t} \hat{\rho} = [\hat{H_0}+\hat{H_I},\hat{\rho}] + \mathcal{L} \hat{\rho} \, ,
	\label{eq:OperatorEquation}
\end{equation}
where $\hat{H_0}$ denotes the unperturbed nuclear Hamilton operator, $\hat{H_I}$ is the interaction Hamilton operator characterizing the effect of the irradiating  laser electric field, and $\mathcal{L}$ represents the Lindblad operator describing the various spontaneous relaxation channels. The matrix elements 
describing the system are denoted by $\rho_{ij}(M_i,M_j)$ with $i,j \in \{e,g\}$ where the respective magnetic sublevels for the ground and excited states are indicated by the magnetic spin quantum numbers in parentheses. We obtain the optical Bloch equations,
\begin{eqnarray}
	\frac{\partial}{\partial t}  \rho_{gg}(M_{g}) 
	&=& - \frac{2}{\hbar} \mathrm{Im}\left(\sum_{M_{e}} \langle I_{e},M_{e}|\hat{H}_{I}|I_{g},M_{g}\rangle e^{+i\omega_{k} t} \rho_{ge}(M_{g},M_{e})\right) \nonumber\\
	& +& \sum_{M_{e}} \gamma(M_{g},M_{e})\rho_{ee}(M_{e}) \, , \nonumber \\
	\frac{\partial}{\partial t}  \rho_{ee}(M_{e}) 
	&=& \frac{2}{\hbar} \mathrm{Im}\left(\sum_{M_{g}} \langle I_{e},M_{e}|\hat{H}_{I}|I_{g},M_{g}\rangle e^{+i\omega_{k} t} \rho_{ge}(M_{g},M_{e})\right) \nonumber\\
	& -& \rho_{ee}(M_{e}) \sum_{M_{g}} \gamma(M_{g},M_{e}) \, , \nonumber \\
	\frac{\partial}{\partial t}  \rho_{ge}(M_{g},M_{e}) 
	&=& i \Delta \rho_{ge} + \frac{i}{\hbar}\langle g|\hat{H}_{I}|e\rangle e^{-i\omega_{k} t} \left( \rho_{gg} - \rho_{ee}\right) \nonumber\\
	& -& \frac{\gamma(M_{g},M_{e})}{2} \rho_{ge}(M_{g},M_{e}) - \gamma_{dec} \rho_{ge}(M_{g},M_{e}) \, ,
	\label{eq:Bloch}
\end{eqnarray}
where $\omega_k$ is the laser frequency, $\Delta=\omega_0-\omega_k$ is the detuning of the laser frequency with respect to the transition frequency, and $\gamma(M_{g},M_{e})$ is the (partial) decay rate corresponding to transitions between definite excited and ground state magnetic sublevels. The partial decay rates are related to the Clebsch-Gordan coefficients $\langle j_{1} m_{1}, j_{2} m_{2}|jm\rangle$ \cite{RingSchuck},
\begin{equation}
	\gamma(M_{g},M_{e}) = \frac{2I_{e}+1}{2L+1} \left[\langle I_{g} M_{g}, I_{e} -M_{e}|LM\rangle\right]^{2} \Gamma_0 \, ,
	\label{eq:PartialDecayRates}
\end{equation}
where $L$ and $M$ denote the photon multipolarity and its total angular momentum projection, respectively, and $\Gamma_0$ is the total decay rate of the excited state. In Eq. (\ref{eq:Bloch}) another decay term $\gamma_{dec}$ that fastens the decay of the off-diagonal elements has been introduced. This stands for the decoherence rate that takes into account a possibly limited coherence time of the laser used in the experiment. The interaction Hamiltonian between the nucleus and the electromagnetic field can be written as 
\begin{eqnarray}
	\hat{H}_{I}=-\frac{1}{c} \int \vec{j}(\vec{r}) \vec{A}(\vec{r},t) d^{3}r \, ,
	\label{eq:InteractionHamiltonian1}
\end{eqnarray}
where $c$ is the speed of light, $\vec{j}(\vec{r})$ denotes the nuclear charge current, and $\vec{A}(\vec{r},t)$ represents the  vector potential of the electromagnetic field. Usually nuclear transitions have a specific multipolarity or present weak multipole mixing. When decomposing the vector potential into spherical waves and additionally assuming only one multipolarity, we obtain for the interaction Hamiltonian matrix element \cite{Adriana_DipoleForbidden}
\begin{eqnarray}
	& \langle I_{e}M_{e}|\hat{H}_{I}|I_{g}M_{g}\rangle  \nonumber\\
	\sim \ & \mathcal{E}_{k} e^{-i\omega_{k}t} \sqrt{2\pi} \sqrt{\frac{L+1}{L}} \frac{k^{L-1}}{(2L+1)!!} \langle I_{e} M_{e}, I_{g} \ -M_{g}|L \ -\sigma \rangle \nonumber\\
	& \times \sqrt{2I_{g}+1} \sqrt{B(\mathcal{\lambda}L,I_{g}\rightarrow I_{e})} \, ,
	\label{eq:InteractionHamiltonian2}
\end{eqnarray}
where $\mathcal{E}_k$ is the electric field amplitude, $k$ denotes the photon wave number, $\sigma$ represents the photon polarization and $B(\mathcal{\lambda}L,I_{g}\rightarrow I_{e})$ is the reduced nuclear transition probability. Eq. (\ref{eq:InteractionHamiltonian2}) is valid for both magnetic and electric multipole transitions. The calculation of the reduced transition probability requires knowledge of the nuclear wave function within a nuclear model. In order to keep a large degree of generality and to be independent of any particular theoretical model it is  more appropriate to take the experimental value of the reduced transition probability, listed for example in online databases such as \cite{NNDC}.
The reduced transition probability is also connected to the radiative decay rate by the expression \cite{RingSchuck}

\begin{eqnarray}
	\Gamma_{rad} = \frac{2L+2}{\epsilon_0 L ((2L+1)!!)^{2}} \left( \frac{E_{\gamma}}{\hbar c} \right)^{2L+1} B(\lambda L,I_{e} \rightarrow I_{g}) \, ,
	\label{eq:RadiativeDecayRate}
\end{eqnarray}
where $\epsilon_0$ is the vacuum permittivity and $E_{\gamma}$ the transition energy.

\subsection{Effective laser parameters} \label{subsec:EffectiveElectricField}
%%%%%%%%%%%%%%%%%%%%%%%%%%%%%%%%%%%%%%%%%%%%%%%%%%%%%%%%%%%%%%%%%%%%%%%%%%%%%%%%%%%%%%%%%%%%%%%%%%%

According to Eq.~(\ref{eq:InteractionHamiltonian2}), the electric field amplitude of the laser light needs to be specified in order to obtain the interaction Hamiltonian matrix element. This quantity has to be related to  the laser parameters typically given in technical design reports of XFELs. The interaction Hamiltonian matrix element depends on the \emph{effective} electric field amplitude, i.e. only photons resonant with the nuclear transition contribute to the laser-nucleus interaction. Nuclear transition widths are typically very small ($10^{-5}...10^{-10} \, \mathrm{eV}$), whereas current XFELs achieve bandwidths in the order of $1 \, \mathrm{eV}$. Accordingly, only a small fraction of all photons within the XFEL pulse meet the resonance condition. To calculate the effective electric field $\mathcal{E}_k$ we need to know the total peak intensity $I_{p}$ of the photon beam and the number of photons from the laser which are resonant with the nuclear transition.\\

The total photon flux $\Phi_{tot}$ can be obtained from the peak power $P_{p}$ and the photon energy $E_{ph}$, $\Phi_{tot} = P_{p}/E_{ph}$. The total number of photons per pulse is related to the pulse duration $T_{p}$ by $N_{tot} = \Phi_{tot} T_{p}$. With the width of the nuclear transition $\Gamma_0$, bandwidth of the laser $BW$, and transition energy $E_{\gamma}$, the resonant photon flux reads
\begin{equation}
	\Phi_{res} = \Phi_{tot} \frac{\Gamma_0}{BW \cdot E_{\gamma}} \, .
\label{phires}
\end{equation}
The number of  resonant photons per pulse is then the product of the resonant photon flux $\Phi_{res}$ and the pulse duration $T_{p}$, $N_{res} = \Phi_{res} T_{p}$. The total peak intensity $I_{p}$ can be determined from the focal spot $A_{foc}$ and peak power $P_{p}$,
\begin{eqnarray}
	 A_{foc} &=& \pi \left( \frac{d_{foc}}{2} \right)^{2} \, , \nonumber \\
	 I_{p} &=& \frac{P_{p}}{A_{foc}} \, ,
\end{eqnarray}
where $d_{foc}$ is the focal diameter. Subsequently, the effective electric field can be calculated from the effective intensity via
\begin{eqnarray}
	 I_{ef}&=& I_{p} \frac{N_{res}}{N_{tot}} \, , \nonumber\\
	 \mathcal{E}_{ef} &=& \sqrt{\frac{2 I_{ef}}{\epsilon_{0} c}} \, .
	\label{eq:EffectiveElectricField}
\end{eqnarray}

As apparent in Eq.~(\ref{phires}), the nuclear transition width $\Gamma_0$ is a crucial quantity that limits the number of resonant photons within the laser bandwidth and 
thus the strength of the field intensity and electric field amplitude. 
 Typically, for neutral atoms, the nuclear width can be written as the sum of the radiative and IC decay rates.
In case of a solid state target,  the collective nature of the nuclear excitation can lead to a broadening of the nuclear bandwidth, as discussed in the following.

%%%%%%%%%%%%%%%%%%%%%%%%%%%%%%%%%%%%%%%%%%%%%%%%%%%%%%%%%%%%%%%%%%
\subsection{Collective effects} \label{subsec:CollectiveEffects}
%%%%%%%%%%%%%%%%%%%%%%%%%%%%%%%%%%%%%%%%%%%%%%%%%%%%%%%%%%%%%%%%%%
New generation lasers like the XFEL have brought with them a tremendous improvement with respect to brightness, coherence and spectral bandwidth in the keV regime. However, nuclear line widths are still orders of magnitude smaller than all that we can achieve with even the best light sources today. Therefore, an important goal concerning laser-nucleus interaction is not only the improvement of laser bandwidths, but one is also seeking for potential mechanisms that could increase the natural line widths of nuclei. 

The possibility to use collective effects for line width enhancement in the interaction of light with a sample of identical nuclei relies on 
the  recoilless absorption or emission of x-ray photons, i.e. the M\"ossbauer effect. For nuclei in a solid state target the photon momentum can be transferred to the crystal lattice as a whole rather than to a single nucleus. Typically recoilless transitions  involve an excited level with lifetimes in the range of 10~ps and energies between  5 and 180 keV. Longer (shorter) lifetimes than indicated lead, according to the Heisenberg uncertainty principle, to too narrow (broad) emission and absorption lines, which no longer effectively overlap. Even for samples of M\"ossbauer nuclei, the probability of recoilless absorption and emission is mostly less than  one, and can be  approximated in the Debye model
\cite{Barb},
\begin{eqnarray}
	f_{LM} = exp \left[ -\frac{2E_R}{k_B \theta_D} \left( 1+4\frac{T^2}{\theta_{D}^{2}} \int_{0}^{\frac{\theta_D}{T}} \frac{xdx}{e^x-1} \right) \right] \, ,
	\label{eq:LambMoessbauerFactor}
\end{eqnarray}
by what is called the Lamb-M\"ossbauer factor $f_{LM}$. In the above expression, $k_B$ is the Boltzmann constant, $\theta_D$ is the Debye temperature, $T$ represents the solid state target temperature, and $E_R=\frac{\hbar^2k^2}{2M}$ denotes the recoil energy. We assume throughout this work that the solid state target is at room temperature $T=300 \, K$, as this does not have a significant effect on our results.\\

A single nucleus that is excited by an XFEL pulse can decay back to the ground state either via radiative decay or IC with a finite constant rate  $\Gamma$. If nuclei are, however, bound inside a crystal lattice,  the excitation generated by the absorption of a photon will not be localized at one single nucleus, but  rather spread out across a large number of nuclei all sharing and contributing to this so-called \emph{excitonic state} \cite{HannonTrammell}. Obviously, this is only possible via coherent photon scattering, meaning that a nucleus absorbing a photon decays back to its initial state upon re-emission. By any incoherent process like IC, nuclear recoil or spin-flip it would be possible to reveal the source's location and therefore tell which nucleus was excited. The decay of the excitonic state occurs not only via the known radiative and IC channels, but also via a time-dependent coherent channel. The coherent decay channel at the time of the  creation of the exciton contributes to the width of the nuclear state and thus in turn to the number of  photons in the laser pulse resonant with the nuclear transition.

\begin{figure}[h]
	
		\includegraphics[width=8cm]{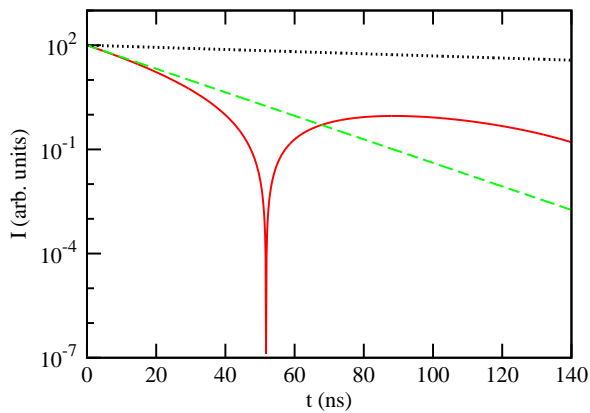}		
	
	\caption{ Intensity of coherently scattered light in Eq.~(\ref{itot}) (solid red line), incoherent natural decay $e^{-\tau}$ (upper dotted black line) and enhanced decay in Eq.~(\ref{isuperrad}) (lower dashed green line) for  the $^{57}\mathrm{Fe}$ isotope (without hyperfine splitting) as a function of time over the duration of one excited nuclear state lifetime. The nuclear sample is assumed to have the
effective thickness $\xi=10$.}
	\label{fig:QuantumBeats}
\end{figure}

There are generally two approaches to deduce the time dependent coherent decay rate. One approach calculates the response function $G(t)$ as done in Ref.~\cite{Kagan} to obtain the time-dependent intensity. The other case \cite{Shvydko_Forward,Motif} is briefly presented in the following. Starting out from Maxwell's equations one  obtains \cite{Shvydko_Forward}
\begin{equation}
	\left( \nabla^2-\frac{1}{c^2} \frac{\partial^2}{\partial t^2} \right) \vec{\mathcal{E}} = \frac{4\pi}{c} \frac{\partial}{\partial t} \vec{\mathcal{J}} \, ,
\end{equation}
where $\vec{\mathcal{E}}$ is the electric field vector of the laser light and $\vec{\mathcal{J}}$ denotes the nuclear source current. In the slowly varying envelope approximation \cite{Scully} and assuming the light propagating in $z$-direction we obtain an equation for the envelopes $\vec{E}$ and  $\vec{J}$,
\begin{equation}
	\frac{\partial}{\partial z} \vec{E} = -\frac{2\pi}{c} \vec{J} \, .
\end{equation}
The electric field amplitude of the radiation re-emitted coherently in forward direction in second-order perturbation theory can be determined via the wave equation \cite{Shvydko_Forward}
\begin{equation}
	\frac{\partial \vec{E}(z,t)}{\partial z} = -\sum_l K_l \vec{J_l}(t) \int_{-\infty}^t \vec{J_l}^{\dagger}(\tau)\cdot \vec{E}(z,\tau) d\tau \, ,
	\label{eq:WaveEquation}
\end{equation}
where $\vec{J_l}(t)$ denotes the nuclear transition current matrix element for the transition specified by the index $l$ that runs over all possible transitions between hyperfine states.  The coefficients $K_l$ characterize the transition and were defined in Ref.~\cite{Shvydko_Forward}. Assuming an initial laser pulse $\vec{E}(t) = \vec{\mathcal{E}_0}\delta(t)$ which is short on the time scale of nuclear dynamics, Eq. (\ref{eq:WaveEquation}) can be solved iteratively, such that  the electric field can be written as a sum
\begin{eqnarray}
	\vec{E}(z,t) = \sum_{n=0}^{\infty} \vec{E}_n(z,t) \, ,
\end{eqnarray}
where each summation term represents a multiple scattering order. The intensity behind the sample is given by $I(t)=|\vec{E}(d,t)|^2$, with  $d$  the target thickness. Assuming only one transition being driven by the laser pulse and disregarding hyperfine splitting, the intensity can be obtained analytically \cite{HannonTrammell,Shvydko_Forward},
\begin{equation}
	I(\tau)=\mathcal{E}_0^2\xi\frac{e^{-\tau}}{\tau}\left[ J_1(\sqrt{4\xi\tau}) \right]^2 \, ,
\label{itot}
\end{equation}
where $\tau=\Gamma_0 t$ is the dimensionless time coordinate and $\Gamma_0$ denotes the transition's total decay rate (radiative + IC). The notation $\xi= \sigma_R N d/4$
is used for the so-called  dimensionless thickness parameter, where 
\begin{eqnarray}
	\sigma_R = 2\pi \frac{2I_e+1}{2I_g+1} \left(\frac{\hbar c}{E_{\gamma}}\right)^2 \frac{1}{1+\alpha} f_{LM}
\end{eqnarray}
is the radiative nuclear resonance cross section and  $N$ is the number density of M\"ossbauer nuclei in the sample.
Furthermore, $\alpha$ denotes the IC coefficient. In the limit of small times $\tau \lesssim 1/\xi$ the Bessel function $J_1$ can be expanded in a Taylor series and we approximately obtain
\begin{equation}
	I(\tau)=\mathcal{E}_0^2\xi^2 e^{-(\xi+1)\tau}
\label{isuperrad}
\label{intensity}
\end{equation}
immediately after excitation. This clearly shows that the coherent decay is faster compared to the spontaneous incoherent $e^{-\tau}$ decay. Fig. \ref{fig:QuantumBeats} illustrates the relation between the different types of decay. The solid red line in Fig. \ref{fig:QuantumBeats}  shows the coherently scattered radiation intensity vs. time for the $^{57}_{26} \mathrm{Fe}$ isotope. The dotted black line in Fig. \ref{fig:QuantumBeats}  indicates the normal $e^{-\tau}$ decay, whereas the dashed green line corresponds to the enhanced $e^{-(\xi+1)\tau}$ decay. For transitions between nuclear levels with hyperfine splitting, the intensity of the scattered light is modulated by the quantum beat due to interference between several unresolved hyperfine transitions. For our numerical analysis, however, the hyperfine splitting can be neglected due to the very short laser pulse length compared to the time scale of the quantum beats. 

From Eq.~(\ref{intensity}) we can also determine the enhancement factor of the decay rate,
\begin{eqnarray}
	\Gamma \simeq (\xi+1) \Gamma_{0} \, ,
	\label{eq:TotalDecayRate}
\end{eqnarray}
where $\Gamma_{0}$ is the spontaneous, isolated nucleus decay rate including the radiative and IC channel and $\Gamma$ additionally includes the enhancement due to collective effects.  Since the effective thickness can take very large values (an effective thickness of $\xi=100$
corresponds to the actual sample thickness of only approx. $d=$20~$\mu$m in the case of  $^{57}\rm{Fe}$), the enhancement factor can be substantial.
However, obviously the effect can not persist for arbitrary large values of the effective thickness $\xi$.  The actual length $l_{c}$ in the sample thickness over which the coherent excitation can occur  is limited by scattering and absorption processes, i.e. the photoelectric effect. Due to these processes the photon beam loses intensity corresponding to a characteristic photo-absorption length $1/\mu$,  unique for every material and usually on the order of several $\mathrm{\mu m}$ up to  tens of $\mathrm{\mu m}$. The effective thickness $\xi$ is therefore limited by the finite length in space over which the x-ray photons penetrate unperturbed inside the solid state target.
 Typically, it is assumed that the length over which the coherent excitation can occur is on the order of the characteristic photo-absorption length. For instance,  the enhancement for the 14.4 keV resonance in $^{57}\mathrm{Fe}$ in NFS geometry is limited by photoabsorption to $10^3$ \cite{Hannon1999}. The effect of absorption in the crystal and of laser focusing (so far we implicitly assumed that the laser beam focus extends over the whole crystal thickness) will be addressed in detail in the next section.

%%%%%%%%%%%%%%%%%%%%%%%%%%%%%%%%%%%%%%%%%
\section{Numerical results} \label{sec:NumericalResults}	
%%%%%%%%%%%%%%%%%%%%%%%%%%%%%%%%%%%%%%%%%

In this section  we present our numerical results of the excited state population after a single radiation pulse  and expected signal photon rate for several nuclear transitions.  We draw a line at transition energies of $E_{\gamma}=25 \, \mathrm{keV}$ and assume that for all nuclei below this value the XFEL  can deliver photons of the necessary energy such that a solid state target can be used. For these cases, we investigate the magnitude of the cooperative effects and the corresponding limitations.  Our choice of 25~keV transition energy is related to the predicted maximum photon energy value for the European XFEL and for XFELO. However, higher harmonics as the ones achieved at LCLS \cite{ratnersecond} or the primary beams of future facilities such as MaRIE  (Matter-Radiation Interactions in Extreme) \cite{MaRIE} may provide photons with energies above 25 keV. We also give at the end of this section an update of previous estimates of laser-nucleus interaction for ion beam targets where the relativistic Doppler effect is used to tune the XFEL photons in resonance with the nuclear transition. For these cases, nuclear transitions with energies up to 100 keV are investigated.

Regardless of the transition energy, we expect the photons to be resonant with the nuclear transition (directly or via relativistic Doppler effect) except for deviations that are small compared to the transition energy, so we can set the detuning $\Delta \simeq 0$. For our case study we use both the actual performance parameters for LCLS or SACLA, the expected values for the XFEL machines still under construction (European XFEL) or envisaged (XFELO), as well as  fully temporally coherent x-ray laser pulses. Our results confirm that coherence is a key ingredient for efficient laser-nucleus interaction.
 As it has been shown in Ref.~\cite{NuclearQuantumOptics}, the coherence sensitivity can be reduced at the expense of XFEL intensity. In the case of the  LCLS, SACLA and  European XFEL  parameters, full coherence accounts to the assumption that these facilities are equipped with a seeding undulator as described in Refs.~\cite{sxfel.Saldin,sxfel.Feldhaus,sxfel.geloni}. Therefore, the energy bandwidth of the laser is given by the Fourier limit and we can assume full temporal coherence  of the radiation pulses resulting in $\gamma_{dec}=0$. If not otherwise specified, we assume an x-ray focal spot diameter of $100 \, \mathrm{nm}$ \cite{EuropeanXFELMachineParameters}. The XFEL parameter data used in the calculations are given in Table~\ref{tab:XFELParameters}.

Upon selecting various isotopes and nuclear transitions we have considered   nuclei that have either a stable or at least long-lived ground state.  All nuclear parameters  were taken from Ref. \cite{NNDC}, while the Debye temperatures $\theta_D$ can be traced back to Refs. \cite{NIST,Hunklinger,Kittel}, and the number  densities of solid targets and x-ray mass attenuation coefficients $\mu/\rho$ to Ref. \cite{NIST}, respectively.

\begin{table}[H]
\caption{ \label{tab:XFELParameters}  Laser beam parameters of LCLS \cite{slac.nature,slac.prl,LCLSMachineParameters}, SACLA \cite{SaclaNatPhot,SaclaMachineParameters2}, European XFEL \cite{EuropeanXFELMachineParameters} and XFELO \cite{XFELO,ydkoprivate}: maximum photon energy $E_{max}$, bandwidth $BW$, pulse duration $T_p$, coherence time $T_{coh}$ (except for SACLA), peak power $P_{peak}$, peak and average brilliance $B$ and pulse repetition rate. The numbers for the European XFEL and XFELO correspond to their expected performance, while the ones for LCLS and SACLA are experimentally confirmed values.}
\begin{indented}	

\item[]		

	\begin{tabular}{lcccc}
\br
	\textbf{Parameter} &  \textbf{LCLS} & \textbf{SACLA} & \textbf{European XFEL} & \textbf{XFELO} \\ \mr
	 $E_{max}$ (eV)  & 10332 & 19556 &  24800 \cite{EuropeanXFELWavelength}  & 25000  \\ 
	 $BW$ &   $2-5\cdot10^{-3}$ & $<10^{-3}$  &   $8\cdot10^{-4}$ & $1.6\cdot10^{-7}$  \\ 
	 $T_{p}$ (fs) &  70--100 & 10  &  100 & 1000  \\ 
	 $T_{coh}$ (fs) & 2  & - & 0.2 &  $1000$ \\ 
	 $P_{peak}$ (W) &  $1.5-4\cdot10^{10}$ & $10^{10}$ & $2\cdot10^{10}$ & $4.1\cdot10^{9}$  \\ 
	 $B_{peak}$$^{*}$ & $2\cdot10^{33}$ & $10^{33}$$^{\dagger}$ & $5.4\cdot10^{33}$ &  $10^{35}$  \\ 
	 $B_{average}$$^{*}$  &  $6\cdot10^{21}$ & $10^{20}$$^{\dagger}$ & $1.6\cdot10^{25}$ & $1.5\cdot10^{27}$ \\
	 Rep. rate (Hz) & $3 \cdot 10^1$ & 10 & $4 \cdot 10^4$ & $ 10^8$ \\
\br
	 
	\end{tabular}

\item[] $^{*}$ The unit of brilliance is photons/(s $\cdot$ mm$^{2}\cdot$ mrad$^{2}\cdot0.1\%$\ BW).
\item[] $^{\dagger}$ Not yet experimentally reported; values from the technical design report \cite{SaclaMachineParameters2}.

\end{indented}

\end{table}

To solve the optical Bloch equations, Eqs. (\ref{eq:Bloch}), we assume as initial conditions that the system is in the ground state and that its population is equally distributed among the corresponding magnetic sublevels. This is well justified because the hyperfine energy splitting is by far lower than the thermal energy at room temperature. We can subsequently numerically calculate the population of each magnetic sublevel of the excited state after one pulse length $T_{p}$ from the Bloch equations (\ref{eq:Bloch}) using the interaction matrix element (\ref{eq:InteractionHamiltonian2}). The nuclear parameters $I_e$, $I_g$, the transition energy and multipolarity and the reduced nuclear transition probability $B(\mathcal{\lambda}L,I_{g}\rightarrow I_{e})$ are taken from nuclear databases such as \cite{NNDC}. The effective field intensity $\mathcal{E}_{ef}$ is estimated using the relevant laser parameters according to the procedure described in Sec. \ref{subsec:EffectiveElectricField}. 
The total excited state population is then the sum over all excited state magnetic sublevels
\begin{eqnarray}
	\rho_{ee} = \sum_{M_e=-I_e}^{I_e} \rho_{ee}(M_e) \, .
\end{eqnarray}
After the radiation pulse, the excited nuclei decay back to their ground state either via the emission of a photon or, when possible an IC electron. The re-emitted photons can be measured in a fluorescence experiment and in first approximation we obtain the signal photon rate
\begin{eqnarray}
	S = \rho_{ee} \ N_{\rm{fv}} \ f_{l} \ \frac{1}{1+\alpha} \, ,
\label{sigrate}
\end{eqnarray}
where $N_{\rm{fv}}$ represents the number of atoms located inside the focal volume of the beam, $f_{l}$ is the pulse repetition rate of the laser and the factor $1/(1+\alpha)$ includes the IC decay channel. For 
bare ions, the IC channel is closed and $\alpha=0$ in the expression above.

%%%%%%%%%%%%%%%%%%%%%%%%%%%%%%%%%%%%%%%%%%%%%%%%%%%%%%%%%%%%%%%%%%%%
\subsection{Solid state targets} \label{sec:NumericsSolidState}
%%%%%%%%%%%%%%%%%%%%%%%%%%%%%%%%%%%%%%%%%%%%%%%%%%%%%%%%%%%%%%%%%%%%
In this section we study several isotopes with transition energies below $25 \, \mathrm{keV}$, namely the cases of $^{57}_{26} \mathrm{Fe}$, $^{73}_{32} \mathrm{Ge}$, $^{83}_{37} \mathrm{Rb}$, $^{119}_{50} \mathrm{Sn}$, $^{134}_{55} \mathrm{Cs}$, $^{137}_{57} \mathrm{La}$, $^{149}_{62} \mathrm{Sm}$, $^{153}_{62} \mathrm{Sm}$, $^{167}_{69} \mathrm{Tm}$, $^{169}_{69} \mathrm{Tm}$, $^{171}_{69} \mathrm{Tm}$, $^{181}_{73} \mathrm{Ta}$, $^{187}_{76} \mathrm{Os}$, $^{193}_{78} \mathrm{Pt}$, $^{201}_{80} \mathrm{Hg}$ and $^{205}_{82} \mathrm{Pb}$. The XFEL pulses shine on  nuclei in a solid state target. We calculate the population in the excited level after a single laser pulse which is resonantly driving the nuclear transition and  the expected signal photon rate.\\

The enhancement of the nuclear width due to collective effects is increasing the number of resonant laser photons and therefore the percentage of excited state population.
The key quantity to be determined is the effective thickness up to which the collective enhancement may occur. Two important parameters have to be taken into account: $(i)$ the laser focal length within which the excitonic state can form and  $(ii)$ the spatial limitation of the excitonic state due to absorption and scattering off electrons which is approximated as the characteristic photo-absorption length. With x-ray focusing in the order of $100 \, \mathrm{nm}$, the focal length is given by twice the Rayleigh length via $L_{foc}=\frac{2\pi}{\lambda} \left(\frac{d_{foc}}{2}\right)^2 \approx 1.6\times10^{-4} \, \mathrm{m}$ for a wavelength of 1~{\AA} \cite{RayleighLength}. 
In Table \ref{tab:NuclearInfo} we present relevant data for nuclear samples on the Lamb-M\"ossbauer factor $f_{LM}$, the characteristic photo absorption length $1/\mu$, the focal length $L_{foc}$ and the dimensionless thickness parameter $\xi$ for $d=1/\mu$.
 We find that the Lamb-M\"ossbauer factor becomes smaller for increasing transition energies. The characteristic photo-absorption length varies between $0.2 \, \mathrm{\mu m}$ and $65 \, \mathrm{\mu m}$ and is always smaller than the corresponding focal length for the considered laser focus of $100 \, \mathrm{nm}$. The dimensionless thickness parameter varies between values close to zero and reaches a maximum of $\xi=87$ for $^{57} \mathrm{Fe}$. The advantages of $^{57}\mathrm{Fe}$ isotope (which is by far the most used M\"ossbauer nucleus so far) become clear with the observation that it has both a large Lamb-M\"ossbauer factor and its crystal lattice allows for considerable effective thickness parameters $\xi$.\\

\begin{table}[h]

\caption{\label{tab:NuclearInfo} For each isotope, the Lamb-M\"ossbauer factor $f_{LM}$, the characteristic photo absorption length $1/\mu$, the focal length $L_{foc}$ for a focus of $1 \, \mathrm{\mu m}$ and the dimensionless thickness parameter $\xi$ for $d=1/\mu$ is presented. The isotopes are ordered by their transition energy.}

\begin{indented}	
		
\item[]	
	\begin{tabular}{cccccc}
	\br
			 Nuclide & E$_{\gamma}$ (keV) & $f_{\rm{ LM}}$ & $\frac{1}{\mu }$  ($\mu $m) & $L_{\rm{ foc}}$ ($\mu $m) & $\xi$  \\ \mr
		$^{201}\mathrm{Hg}$	  & 1.565 & 0.98 & 0.341 & 19.8 & 0.00359 \\
		$^{193}\mathrm{Pt}$	  & 1.642 & 1. & 0.239 & 20.8 & 0.323 \\
		$^{205}\mathrm{Pb}$	  & 2.329 & 0.95 & 1.12 & 29.5 & $2.98\times 10^{-7}$ \\
		$^{151}\mathrm{Sm}$	  & 4.821 & 0.92 & 3.96 & 61.1 & 0.218 \\
		$^{171}\mathrm{Tm}$	  & 5.036 & 0.95 & 1.96 & 63.8 & 0.249 \\
		$^{83}\mathrm{Rb}$	  & 5.260 & 0.2 & 18.3 & 66.6 & 0.556 \\
		$^{181}\mathrm{Ta}$	  & 6.238 & 0.94 & 1.79 & 79. & 2.53 \\
		$^{169}\mathrm{Tm}$	  & 8.410 & 0.85 & 7.56 & 107. & 1.66 \\
		$^{187}\mathrm{Os}$	  & 9.756 & 0.95 & 4.33 & 124. & 1.34 \\
		$^{167}\mathrm{Tm}$	  & 10.400 & 0.78 & 2.9 & 132. & 0.155 \\
		$^{137}\mathrm{La}$	  & 10.560 & 0.5 & 7.23 & 134. & 0.384 \\
		$^{134}\mathrm{Cs}$	  & 11.244 & 0.00015 & 29.3 & 142. & 0.000239 \\
		$^{73}\mathrm{Ge}$	  & 13.284 & 0.75 & 12.5 & 168. & 0.0764 \\
		$^{57}\mathrm{Fe}$	  & 14.413 & 0.76 & 21.9 & 183. & 86.7 \\
		$^{149}\mathrm{Sm}$	  & 22.507 & 0.16 & 33.7 & 285. & 0.472 \\
		$^{119}\mathrm{Sn}$	  & 23.871 & 0.082 & 64.6 & 302. & 6.86 \\
		\br
		\end{tabular}
\end{indented}
\end{table}

Considering the case of $^{57}\mathrm{Fe}$, we proceed to investigate first the magnitude of collective effects for the excited state population. The  M\"ossbauer transition of iron has a large Lamb-M\"ossbauer factor ($f_{\rm{ LM}}\simeq$0.8). 
We study the dependence of the excited state population $\rho_{ee}$ on the focal diameter $d_{foc}$ considering a constant number of XFEL photons in the beam focus. A larger focus  is thus related to the counteracting effects of lower intensity and larger collective nuclear width enhancement. In turn, a smaller focus limits the possible contribution of collective effects on the nuclear transition width but simultaneously  allows for higher laser intensity. We use here for exemplification the XFELO laser parameters.

\begin{figure}[h]
	\includegraphics[width=10cm,angle=0]{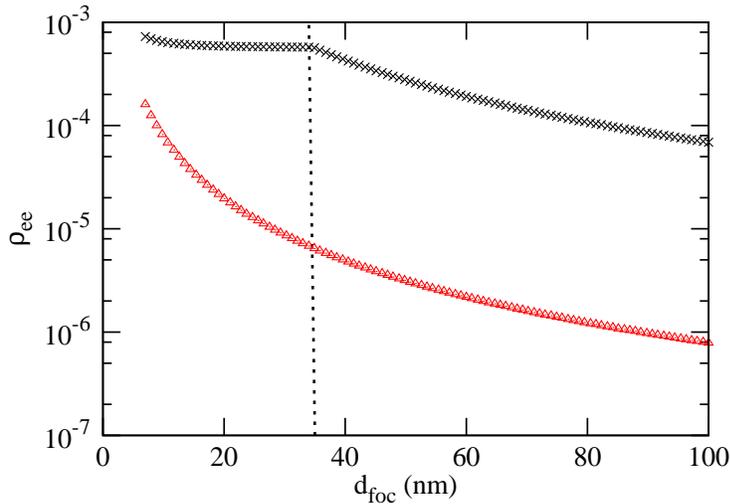}
	\caption{Excited state population $\rho_{ee}$ of $^{57} \mathrm{Fe}$ vs. focal diameter $d_{foc}$ with (black crosses) and without (red triangles) cooperative effects taken into account.   The dashed vertical line indicates the focal diameter where $d=L_{foc}=1/\mu$.}
	\label{fig:NumericsIron}
\end{figure}

Fig. \ref{fig:NumericsIron} shows the excited state population $\rho_{ee}$ as a function of the focal diameter $d_{foc}$ on a logarithmic scale for the $^{57} \mathrm{Fe}$ isotope. We cover focal diameters from $7 \, \mathrm{nm}$ (best focus achieved in Ref. \cite{XRayFocus}) up to $100 \, \mathrm{nm}$.  The sample thickness $d$ is limited on the upper side due to both photo absorption and beam focusing, i.e. we always need to choose the smaller one of the two. For a better visualization, in Fig. \ref{fig:NumericsIron} the $d_{foc}$-axis is separated into two regions by a dashed vertical line, corresponding to the focal diameter at which the focal length is equal to the characteristic photo absorption length of iron ($L_{foc}=1/\mu$).
As a comparison, we also present calculated data for the nuclear excited state population when no collective effects would occur. 

According to Fig. \ref{fig:NumericsIron} we may distinguish two regimes,

\begin{itemize}
	\item \textbf{$d=L_{foc}<\frac{1}{\mu}$}\\
	the sample thickness $d$ is limited on the upper side by the focal length. According to its definition, the dimensionless thickness parameter $\xi$ depends linearly on the sample thickness $d=L_{foc}$, and the focal length $L_{foc}$ depends on the square of the focal diameter: $\xi \propto d_{foc}^2$. The dataset including collective effects (black crosses) is hardly affected by an increase of $d_{foc}$, because the line width enhancement due to the coherent decay compensates for the decreasing photon flux. For the case without collective effects (red triangles), $\rho_{ee}$ decreases significantly on the same interval. We find that at the best focus both datasets differ by a factor of $4.5$, while at the vertical dashed line ($d=L_{foc}=1/\mu$) the discrepancy already reaches a value of $88$. This reflects the fact that at very small focal diameters not many nuclei are located inside the focal volume and thus collective effects become less  important. For all isotopes  for small excited state populations (as in our case) the cooperative effect enhancement factor for $\rho_{ee}$ is approximately $\xi+1$, see Eq. (\ref{eq:TotalDecayRate}).
	\item \textbf{$d=\frac{1}{\mu}<L_{foc}$}\\
	 the sample thickness $d$ is limited on the upper side by the (constant) characteristic photo absorption length, which is about 22~$\mu$m for the $^{57} \mathrm{Fe}$ isotope. In this case, we obtain a constant dimensionless thickness parameter of $\xi \approx 87$. Hence, collective effects do not increase in magnitude any further when proceeding to larger focal diameters and the corresponding dataset (black crosses) decreases parallel to the dataset without collective effects to the right of the vertical dashed line. We can thus deduce the maximum collective effect enhancement factor on the excited state population $\rho_{ee}$ which is limited due to the characteristic photo absorption length. In the case of $^{57} \mathrm{Fe}$ this factor is approximately $\xi(d=1/\mu)+1 \approx 88$.
\end{itemize}

The most interesting differences between the cases with and without collective effects occur at very small focal diameters in the order of several $\mathrm{nm}$. In Ref. \cite{XRayFocus}, a reduction of the x-ray beam focusing up to diameters in the order of $7 \, \mathrm{nm}$ was reported. However, in this case such a tiny focus goes along with a tremendous loss in radiation intensity, i.e., only a small part of the photons in the pulse are focused while the rest are lost. At present, a 100~nm focus spot size achieved with Kirkpack-Baez mirrors is expected to be possible for 95$\%$ of the photons in the pulse \cite{EuropeanXFELMachineParameters}. As an alternative, diffractive lenses for x-rays \cite{SchroerPRL,SchroerAPL,KangAPL} should be able to focus 5$\times 10^{10}$ photons on a 80$\times$80 nm$^2$ focal spot. For our calculations in the following we have assumed  that the full photon number remains in the beam focus and considered the more modest value of 100~nm for the focal spot size.

Taking into account the laser  parameters of several XFEL facilities and  a 100~nm focus spot size, we have calculated the population in the excited level after a single laser pulse $\rho_{ee}$ and the expected signal photon rate $S$ for the considered isotopes with transition energies below 25 keV. The collective enhancement of the nuclear transition width has been taken into account. The results are presented in Tables \ref{tab:SolidState1} (full temporal coherence) and \ref{tab:SolidState2} (presently available laser parameters without seeding). For the case of full temporal coherence we have considered  $\gamma_{dec}=0$ and the Fourier bandwidth of the laser pulse instead of the values given in Table \ref{tab:XFELParameters}. For the case of SACLA, where the coherence time performance has not been reported yet, we have considered the coherence time to be equal to the pulse duration.
The difference between the results with partial and full temporal coherence account for almost six orders of magnitude for LCLS and European XFEL parameters and confirm the crucial importance of the longitudinal coherence of the pulses. The difference in the values of the excited state population  $\rho_{ee}$ and the signal rate $S$  is smaller for SACLA because we have assumed in the first place a long coherence time for the no-seeding case, which is probably optimistic. The SACLA results are overall smaller than the ones for the other parameter sets mostly due to a very short pulse (10~fs) and the presently small repetition rate of 10 Hz. The high repetition rate of the future European XFEL, on the other hand, makes this facility particularly attractive for high signal rates. Seeding simulations have been already performed for this facility, unfortunately mostly however aiming at a shorter pulse duration (7~fs) \cite{sxfel.geloni}. Optimal for nuclear quantum optics applications are however longer pulses, as may be available someday at a future XFELO facility.

\begin{figure}[h]
	\hspace{-1cm}
	\begin{minipage}[t]{8cm}
		\vspace{0cm}
		\includegraphics[width=9cm,angle=0]{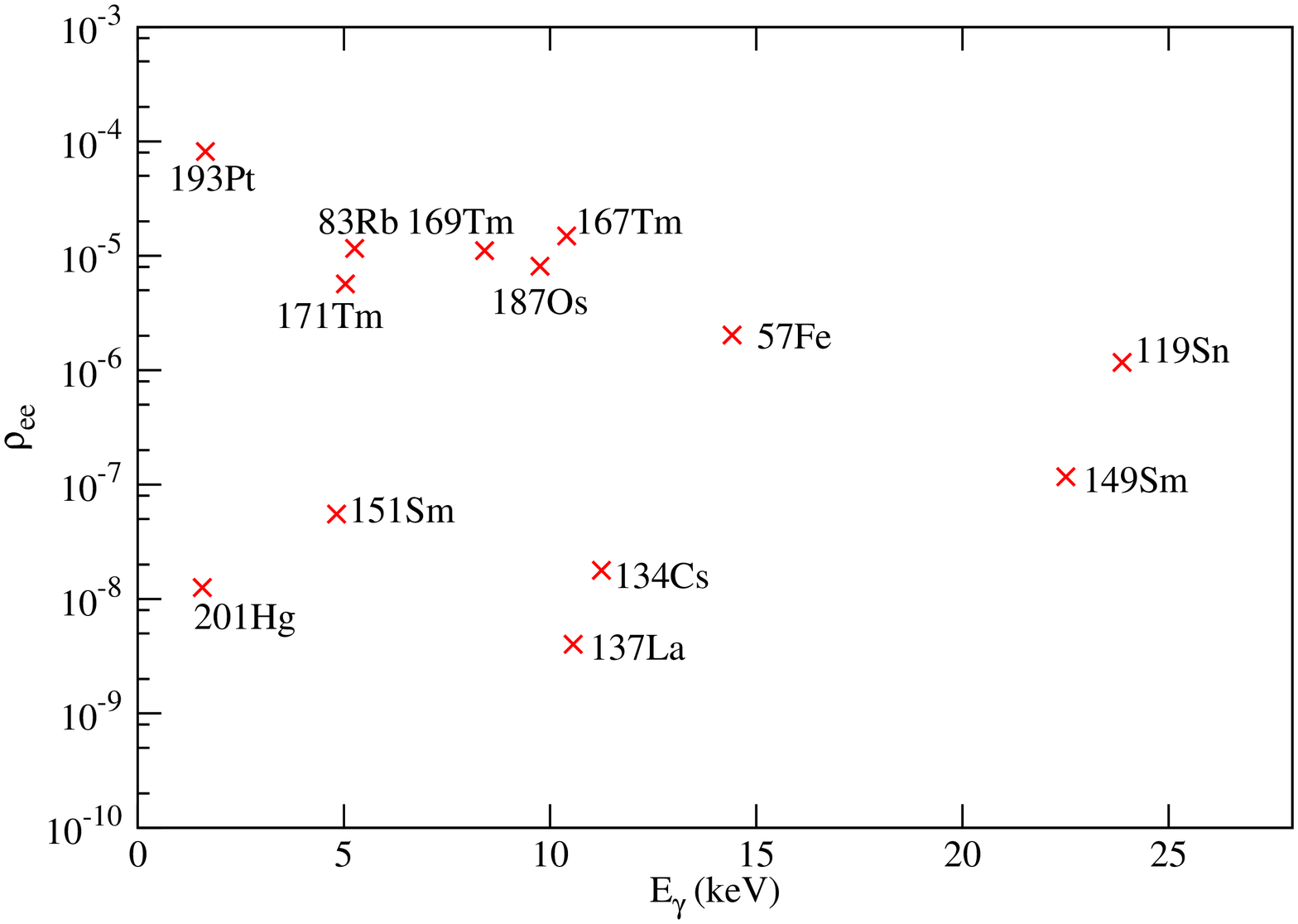}
	\end{minipage}
	%\hfill
	\hspace{1cm}
	\begin{minipage}[t]{8cm}
		\vspace{0pt}
		\includegraphics[width=9cm,angle=0]{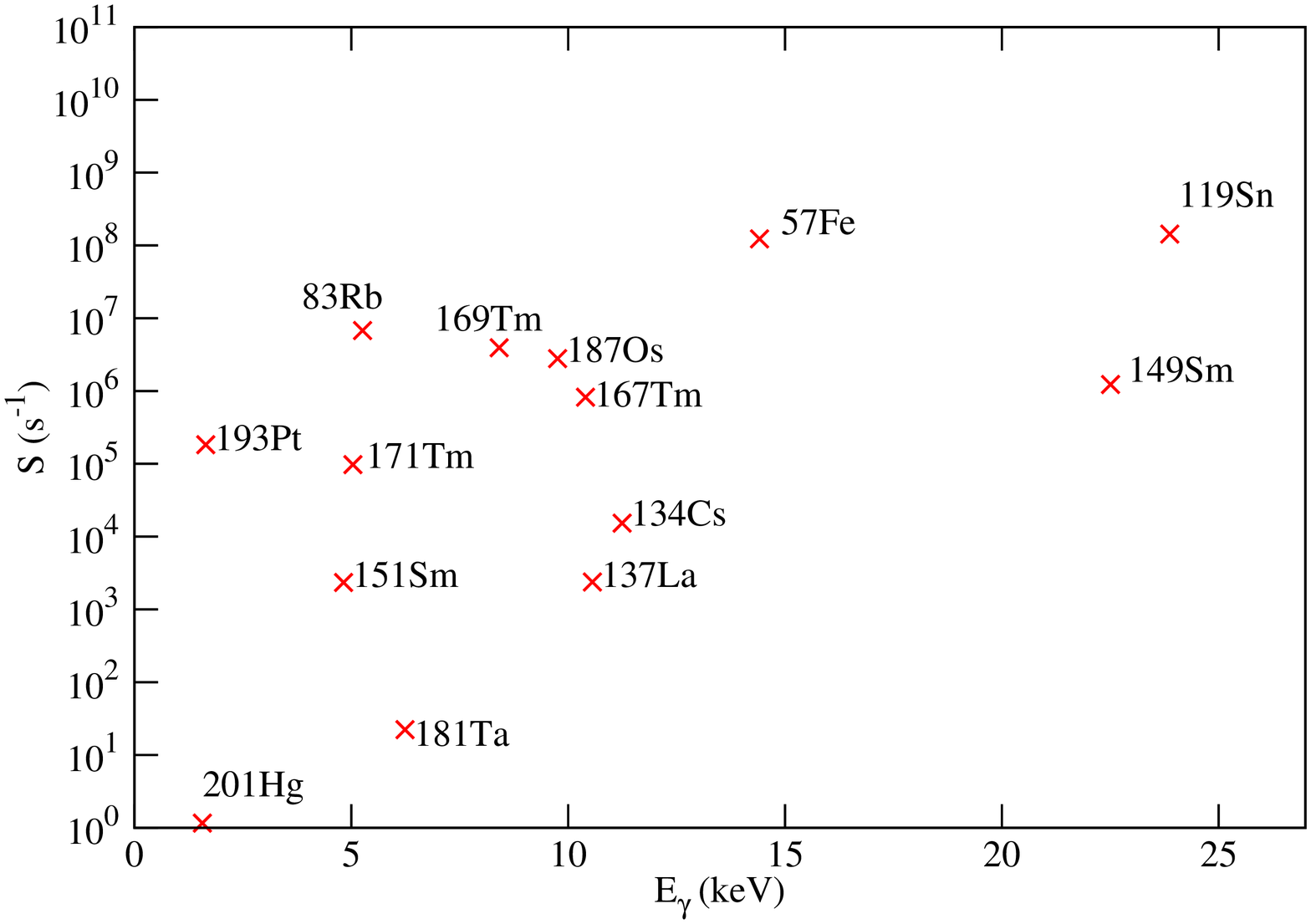}
	\end{minipage}
	%\setcaptionwidth{15cm}
	\caption{Excited state population $\rho_{ee}$ (left) and signal photon rate $S$ (right) vs. transition energy $E_{\gamma}$ obtained for the European XFEL parameters considering full temporal coherence for the isotopes in  Table \ref{tab:SolidState1}.}\label{fig:NumericsRhoSSolidState}
\end{figure}

For a better visualization, we plot the excited state population and the signal rate 
for the European XFEL  parameters with seeding and a pulse duration of 100~fs in 
 Fig.~\ref{fig:NumericsRhoSSolidState}. We refrain from plotting the same figures for SACLA, LCLS and XFELO since all points only have a constant offset factor that can be looked up in Tables  \ref{tab:SolidState1} and \ref{tab:SolidState2}.
 The highest excited state populations are obtained for the $^{193}_{78} \mathrm{Pt}$ isotope:  $\rho_{ee}=8.15\times10^{-5}$ (European XFEL), $\rho_{ee}=1.64\times10^{-4}$ (LCLS), $\rho_{ee}=2.77\times10^{-3}$ (XFELO) and $\rho_{ee}=4.11\times10^{-8}$ (SACLA). The highest signal photon rates are obtained for the $^{119}_{50} \mathrm{Sn}$ isotope, namely  $1.43 \times 10^8 \, s^{-1}$ (European XFEL), $S=2.16 \times 10^5 \, s^{-1}$ (LCLS), $S=1.30\times10^{12} \, s^{-1}$ (XFELO), and $S=1.93\times10^{2} \, s^{-1}$ (SACLA).  Since LCLS has a lower pulse repetition rate than the European XFEL, the signal photon rates are smaller even though the excited state population obtained with the LCLS parameters is actually higher.
Were the total number of photons per pulse to be successfully focused on  the 7~nm focal spot \cite{XRayFocus}, the achieved intensity would allow for an excited state population as high as $\rho_{ee}=1$ for XFELO laser parameters. One may therefore speculate that the experimental realization of nuclear Rabi-oscillations is related to the future improvement of x-ray optics devices. 

\begin{table}[h]

\caption{\label{tab:SolidState1} Excited state population $\rho_{ee}$ and signal photon rate $S$ for a sample of nuclei with transition energies below $25 \, \mathrm{keV}$ are presented for four sets of laser parameters. Seeding, i.e.,  perfect coherence of the pulse, are considered for the European XFEL, LCLS and SACLA. XFELO is designed from the start to provide fully coherent pulses. The horizontal lines in the LCLS/SACLA column indicate the border between currently accessible photon energies and isotopes with higher-lying first excited states. The numbers in brackets denote the power of ten to  multiply with.}
	
\scriptsize
	
%	\begin{indented}	
		
%\item[]		
		\begin{tabular}{|lrc|ll|ll|ll|ll|}
		 \br
 		 	& $E_{\gamma}\ \, $	 & 	 &  \multicolumn{2}{|c|}{European XFEL} & \multicolumn{2}{|c|}{ LCLS} & \multicolumn{2}{|c|}{ SACLA} & \multicolumn{2}{|c|}{ XFELO} \\ 
Nuclide & \rm{ (keV)} &  $\lambda $L & $\rho_{ee}$ & S ($s^{-1}$) & $\rho_{ee}$ &  S ($s^{-1}$) & $\rho_{ee}$ &  S ($s^{-1}$) & $\rho_{ee}$ &  S ($s^{-1}$) \\ 
\mr
$^{201}\mathrm{Hg}$ &  1.565 & M1 & 1.26(-8) & 1.16 &   2.54(-8) & 1.75(-3) & 6.35(-12) & 1.46(-7) & 4.28(-7) & 9.86(2) \\
$^{193}\mathrm{Pt}$ &  1.642 & M1 & 8.15(-5) & 1.83(5) &   1.64(-4) & 2.76(2) & 4.11(-8) & 2.3(-2) & 2.77(-3) & 1.55(8) \\
$^{205}\mathrm{Pb}$ &  2.329 & E2 & 2.04(-18) & 5.34(-14) & 4.11(-18) & 8.08(-17) & 1.03(-21) & 6.73(-21) & 6.93(-17) & 4.54(-11)  \\
$^{151}\mathrm{Sm}$ & 4.821 & M1 &  5.53(-8) & 2.34(3) & 1.11(-7) & 3.54 & 2.83(-11) & 3.0(-4) & 1.91(-6) & 2.02(6) \\
$^{171}\mathrm{Tm}$ & 5.036 & M1 & 5.68(-6) & 9.76(4) & 1.15(-5) & 1.48(2) & 2.9(-9) & 1.24(-2) & 1.95(-4) & 8.4(7) \\
$^{83}\mathrm{Rb}$ &  5.260 & M1 & 1.16(-5) & 6.76(6) &   2.34(-5) & 1.02(4) & 1.4(-8) & 2.04 & 9.47(-4) & 1.38(10) \\
$^{181}\mathrm{Ta}$  & 6.238 & E1 & 5.15(-11) & 2.23(1) &   1.04(-10) & 3.37(-2) & 2.72(-14) & 2.94(-6) & 1.84(-9) & 1.98(4) \\
$^{169}\mathrm{Tm}$  &  8.410 & M1 &1.11(-5) & 3.93(6) &   2.24(-5) & 5.94(3) & 6.2(-9) & 5.48(-1) & 4.18(-4) & 3.69(9) \\
$^{187}\mathrm{Os}$  &  9.756 & M1 &  8.1(-6) & 2.79(6) &   1.63(-5) & 4.22(3) & 4.2(-9) & 3.62(-1) & 2.83(-4) & 2.44(9) \\ \cline{6-7} 
$^{167}\mathrm{Tm}$ & 10.400 & M1 & 1.49(-5) & 8.21(5) &   3.01(-5) & 1.24(3) & 7.81(-9) & 1.07(-1) & 5.27(-4) & 7.24(8) \\
$^{137}\mathrm{La}$  & 10.560 & M1 & 4.02(-9) & 2.38(3) &   8.1(-9) & 3.6 & 2.6(-12) & 3.84(-4) & 1.75(-7) & 2.59(6) \\
$^{134}\mathrm{Cs}$  & 11.244 & M1 & 1.78(-8) & 1.53(4) & 3.6(-8) &   2.32(1) & 2.35(-11) & 5.06(-3) & 1.59(-6) & 3.41(7) \\
$^{73}\mathrm{Ge}$ &  13.284 & E2 & 1.01(-13) & 1.56(-2) &   2.03(-13) & 2.35(-5) & 5.21(-17) & 2.01(-9) & 3.51(-12) & 1.35(1) \\
$^{57}\mathrm{Fe}$ &  14.413 & M1 &  2.03(-6) & 1.23(8) &   4.08(-6) & 1.86(5) & 1.34(-9) & 2.04(1) & 9.03(-5) & 1.37(11) \\ \cline{8-9}  
$^{149}\mathrm{Sm}$ & 22.507 & M1 &  1.17(-7) & 1.23(6) &   2.35(-7) & 1.85(3) & 1.61(-10) & 4.23(-1) & 1.09(-5) & 2.85(9) \\
$^{119}\mathrm{Sn}$ & 23.871 & M1 & 1.17(-6) & 1.43(8) &   2.35(-6) & 2.16(5) & 6.33(-9) & 1.93(2) & 4.27(-4) & 1.3(12) \\
\br
\end{tabular}
%\end{indented}	
	
\normalsize
	\end{table}

\begin{table}[h]

\caption{\label{tab:SolidState2} Excited state population $\rho_{ee}$ and signal photon rate $S$ this time without seeding, using the actual performance parameters for LCLS and SACLA and the present design values for the European XFEL. The horizontal lines in the LCLS/SACLA column indicate the border between currently accessible photon energies and isotopes with higher-lying first excited states. The numbers in brackets denote the power of ten to  multiply with.}
	
\scriptsize
	
	\begin{indented}	
		
\item[]		
		\begin{tabular}{|lrc|ll|ll|ll|}
		 \br
 		 	& $E_{\gamma}\ \, $	 & 	 &  \multicolumn{2}{|c|}{European XFEL} & \multicolumn{2}{|c|}{ LCLS} & \multicolumn{2}{|c|}{ SACLA} \\ 
Nuclide & \rm{ (keV)} &  $\lambda $L & $\rho_{ee}$ & S ($s^{-1}$) & $\rho_{ee}$ &  S ($s^{-1}$) & $\rho_{ee}$ &  S ($s^{-1}$)  \\ 
\mr

$^{201}\mathrm{Hg}$ &  1.565 & M1 & 1.67(-14) &
   1.54(-6) & 9.09(-14) & 6.28(-9) & 5.58(-15) & 1.29(-10) \\

$^{193}\mathrm{Pt}$ &  1.642 & M1 & 1.08(-10) & 2.42(-1)
   & 5.88(-10) & 9.89(-4) & 3.62(-11) & 2.03(-5) \\

$^{205}\mathrm{Pb}$ &  2.329 & E2 & 2.7(-24) & 7.08(-20) & 1.47(-23) & 2.9(-22) & 9.54(-25) & 6.26(-24) \\

$^{151}\mathrm{Sm}$ & 4.821 & M1 &  7.32(-14) & 3.1(-3) &
   3.99(-13) & 1.27(-5) & 2.49(-14) & 2.64(-7) \\

$^{171}\mathrm{Tm}$ & 5.036 & M1 & 7.52(-12) & 1.29(-1)
   & 4.1(-11) & 5.29(-4) & 2.55(-12) & 1.1(-5) \\

$^{83}\mathrm{Rb}$ &  5.260 & M1 & 1.54(-11) & 8.95 & 8.4(-11) & 3.66(-2) & 1.24(-11) & 1.8(-3) \\

$^{181}\mathrm{Ta}$  & 6.238 & E1 & 6.82(-17) & 2.95(-5) &
   3.72(-16) & 1.21(-7) & 2.42(-17) & 2.61(-9) \\

$^{169}\mathrm{Tm}$  &  8.410 & M1 &1.47(-11) & 5.2 & 8.02(-11) & 2.13(-2) & 5.46(-12) & 4.82(-4) \\

$^{187}\mathrm{Os}$  &  9.756 & M1 &  1.07(-11) & 3.7 & 5.85(-11) & 1.51(-2) & 3.7(-12) & 3.18(-4) \\ \cline{6-7} 

$^{167}\mathrm{Tm}$ & 10.400 & M1 & 1.98(-11) & 1.09 & 1.08(-10) & 4.44(-3) & 6.88(-12) & 9.45(-5) \\

$^{137}\mathrm{La}$  & 10.560 & M1 &  5.32(-15) & 3.15(-3) &
   2.9(-14) & 1.29(-5) & 2.29(-15) & 3.38(-7) \\

$^{134}\mathrm{Cs}$  & 11.244 & M1 &  2.36(-14) & 2.03(-2) & 1.29(-13) & 8.3(-5) & 2.07(-14) & 4.45(-6) \\

$^{73}\mathrm{Ge}$ &  13.284 & E2 &  1.34(-19) & 2.06(-8) &
   7.29(-19) & 8.42(-11) & 4.63(-20) & 1.78(-12) \\

$^{57}\mathrm{Fe}$ &  14.413 & M1 &  2.68(-12) & 1.63(2) &
   1.46(-11) & 6.67(-1) & 1.18(-12) & 1.79(-2) \\ \cline{8-9} 
 
$^{149}\mathrm{Sm}$ & 22.507 & M1 &  1.54(-13) & 1.62 & 8.42(-13) & 6.64(-3) & 1.42(-13) & 3.72(-4) \\

$^{119}\mathrm{Sn}$ & 23.871 & M1 &  1.55(-12) & 1.89(2) &
   8.43(-12) & 7.73(-1) & 5.57(-12) & 1.7(-1) \\
\br
\end{tabular}
\end{indented}	
	
\normalsize
	\end{table}

Finally, we conclude this part with two important observations regarding the approximations used in our calculations and possible experimental issues. Firstly, the formation of the exciton and its coherent decay perturb both the time signature of the spectrum as well as the ratio of emitted photons to emitted IC electrons. In this respect, the signal rate expression in Eq.~(\ref{sigrate}) is a rather poor approximation. Following the iteration procedure described in Sec.~\ref{subsec:CollectiveEffects}, the time-dependent collective width of the nuclear excited states should be accurately calculated \cite{Adriana_MagneticSwitching} (also taking into account the coherence of the laser) and correspondingly the  radiative nuclear decay  expected in fluorescence experiments. 

The second observation is meant to counter the effects of the first one and relates to the ablation or melting of the target. Due to the very strong intensities of the XFEL, far larger than previous available values in SR experiments, solid-state samples may be completely depleted of electrons and, as a consequence of Coulomb repulsion, explode on a time scale shorter than that of the nuclear decay.  In consequence, the decay of the excited nuclei proceeds from a state very much different from the original target. While studies on target damage and nuclear x-ray fluorescence have not been performed so far, we expect that $(i)$ through Coulomb explosion the excitonic state is destroyed and no coherent decay is present, $(ii)$ the nuclear decay proceeds with the spontaneous decay rate known from single nuclei. Presumably the IC channel is restored by  electron recombination into the highly charged ion on a time scale faster than the one of the nuclear decay. The destruction of the sample would make impossible multiple exposure of the same nuclei to the XFEL pulse, but successive burns on a rotating or tape station system solid state nuclear target offer a scenario for the laser-nuclei interaction experiment. The expression for the signal photon rate (\ref{sigrate}) is therefore suited to describe this situation.

Nuclear fluorescence experiments similar to NFS ones can be performed with solid state targets. The main requirement for the targets is to have a high Lamb-M\"ossbauer factor $f_{LM}$, i.e., a large fraction of recoilless nuclear transitions occurring in the sample. This can be achieved either in pure crystals or polycrystalline foils as the ones used in nuclear lighthouse effect experiments \cite{nle}. In principle, the nuclei of interest can also be implanted as dopants in a hard crystalline host with conveniently high Debye temperature, in order to increase the value of $f_{LM}$, as it was suggested in Ref.~\cite{Adriana_MagneticSwitching}. However, estimating the recoilless fraction
of absorption and emission in nuclear transitions for impurities in hard crystalline host materials  requires dedicated calculations. In contrast to NFS experiments with SR, the sample is likely to be destroyed by the pulse, such that the fluorescence photons from the coherent decay will no longer be emitted only in the forward direction. Furthermore, one may expect a high background from the electronic plasma created by the x-ray pulse. Time gating techniques as used in NFS experiments \cite{Hastings1991} may prove themselves also here useful. A good signal-to-background ratio is in this case crucial in order to experimentally confirm for the first time the direct laser-nucleus interaction.

\subsection{Accelerated nuclear targets} \label{sec:NumericsAccelerated}
%%%%%%%%%%%%%%%%%%%%%%%%%%%%%%%%%%%%%%%%%%%%%%%%%%%%%%%%%%%%%%%%%%%%%%%%%%%%%

Here we provide an update of previous values \cite{Adriana_DipoleForbidden,Adriana_Photoexcitation} for the case of x-ray pulses attempting to drive nuclear transitions with higher excitation energies than available directly from the laser. To bridge the gap between photon and nuclear transition frequency, target acceleration has been proposed \cite{NuclearQuantumOptics}. The target nuclei can be accelerated to relativistic speeds towards the photon beam. Due to the relativistic Doppler shift the photon energy becomes higher in the nuclear rest frame than it is in the laboratory frame. Hence, one accelerates the nuclei to a specific velocity $v$ relative to the laboratory system to achieve an overlap between photon energy and transition energy in the nuclear rest frame (subscript $n$), $\omega_{n} = \omega \sqrt{(1+\beta)/(1-\beta)}$, where $\beta$ is the relativistic factor $\beta=v/c$. Here,  $\omega_{n}$ and $\omega$ are the laser frequencies in the nuclear rest frame and in the laboratory frame, respectively. The electric field amplitude of the laser is transformed as $\mathcal{E}_n=\mathcal{E}(1+\beta)\gamma$, where $\gamma = 1/\sqrt{1-\beta^{2}}$. Furthermore, both the natural bandwidth of the laser $BW$ and the relative uncertainty in the gamma factor $\Delta\gamma / \gamma$ influence the bandwidth of the laser in the nuclear rest frame. We assume that both photon energy in the laboratory frame and the gamma factor follow a Gaussian distribution, such that the bandwidth transformed into the nuclear rest frame can be determined by Gaussian error propagation and is approximately given by
\begin{eqnarray}
	BW_n \simeq \sqrt{ \left( BW \right)^{2} + \left( \frac{\Delta\gamma}{\gamma} \right)^{2} } \, .
	\label{eq:Bandwidth}
\end{eqnarray}
The transformed laser bandwidth then enters the calculation of the flux of resonant photons in Eq.~(\ref{phires}).

We analyze the cases of $^{131}_{54} \mathrm{Xe}$, $^{153}_{62} \mathrm{Sm}$, $^{153}_{63} \mathrm{Eu}$, $^{160}_{65} \mathrm{Tb}$, $^{165}_{67} \mathrm{Ho}$, $^{173}_{70} \mathrm{Yb}$, $^{183}_{74} \mathrm{W}$, $^{192}_{77} \mathrm{Ir}$, $^{223}_{88} \mathrm{Ra}$, and $^{195}_{78} \mathrm{Pt}$ having transition energies up to $100 \, \mathrm{keV}$. The maximum photon energy of LCLS is no longer assumed to be $25 \, \mathrm{keV}$ as in Sec. \ref{sec:NumericsSolidState}, but rather $10.3\, \mathrm{keV}$ as stated in Table \ref{tab:XFELParameters} corresponding to photon energies available at present. Similarly, we have considered 19.5 keV photon energy for the SACLA XFEL.
 As in Refs.~\cite{Adriana_DipoleForbidden,Adriana_Photoexcitation} we consider here perfect temporal coherence of the laser pulses, which corresponds to XFEL seeding. 
 The target consists now of bare nuclei without surrounding electrons so that neither collective effects nor IC play a role and only the radiative decay rate $\Gamma_{rad}$ (\ref{eq:RadiativeDecayRate}) is taken into account. In the few cases where the laser-driven nuclear transition does not connect the ground state to the first excited state but to a higher level, we neglect the weak non-resonant coupling between the x-ray pulse and the intermediate lower-lying nuclear states, which  remain unpopulated.

\begin{figure}[h]
	\hspace{-1cm}
	\begin{minipage}[t]{8cm}
		\vspace{0cm}
		\includegraphics[width=9cm,angle=0]{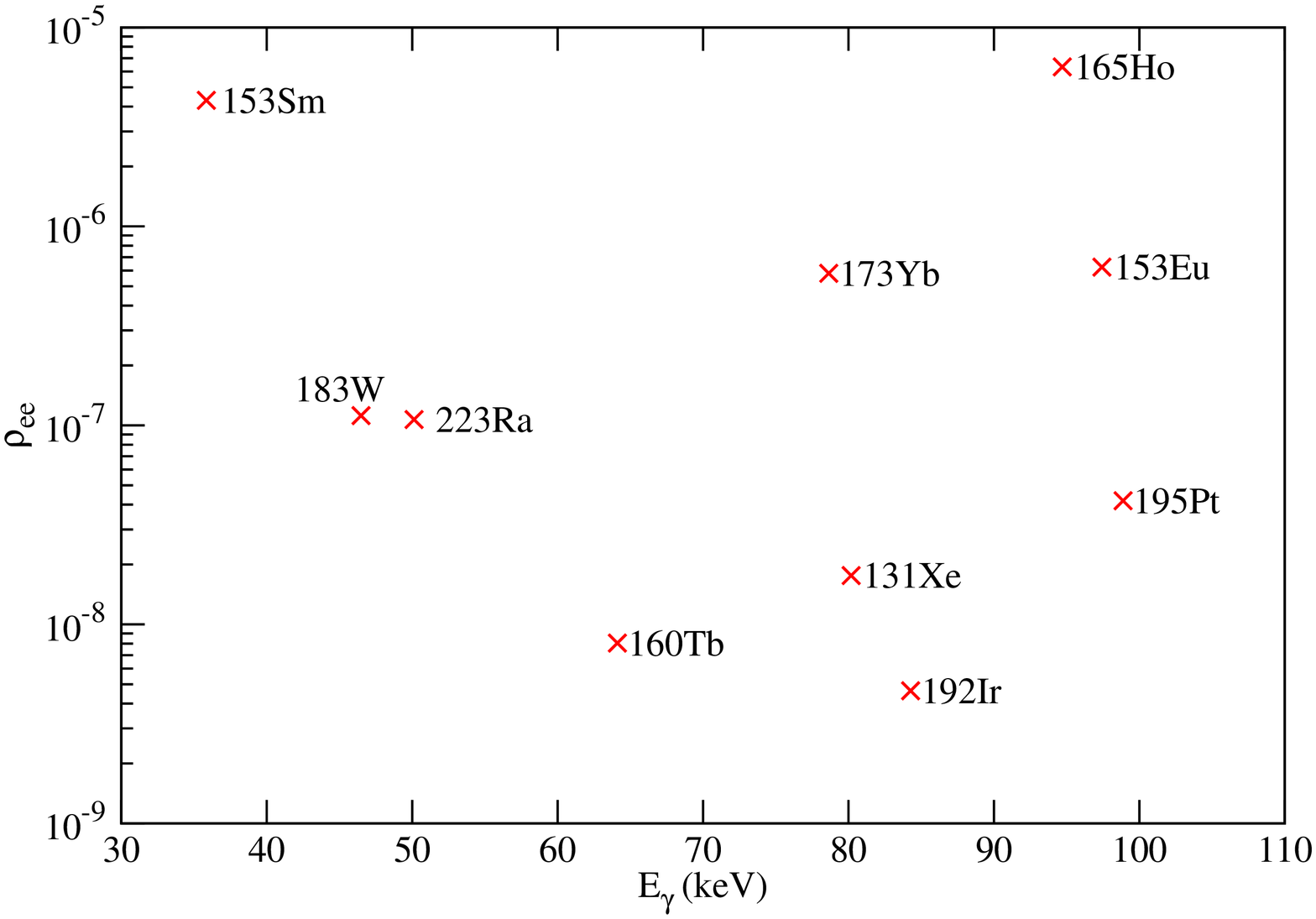}
	\end{minipage}
	%\hfill
	\hspace{1cm}
	\begin{minipage}[t]{8cm}
		\vspace{0pt}
		\includegraphics[width=9cm,angle=0]{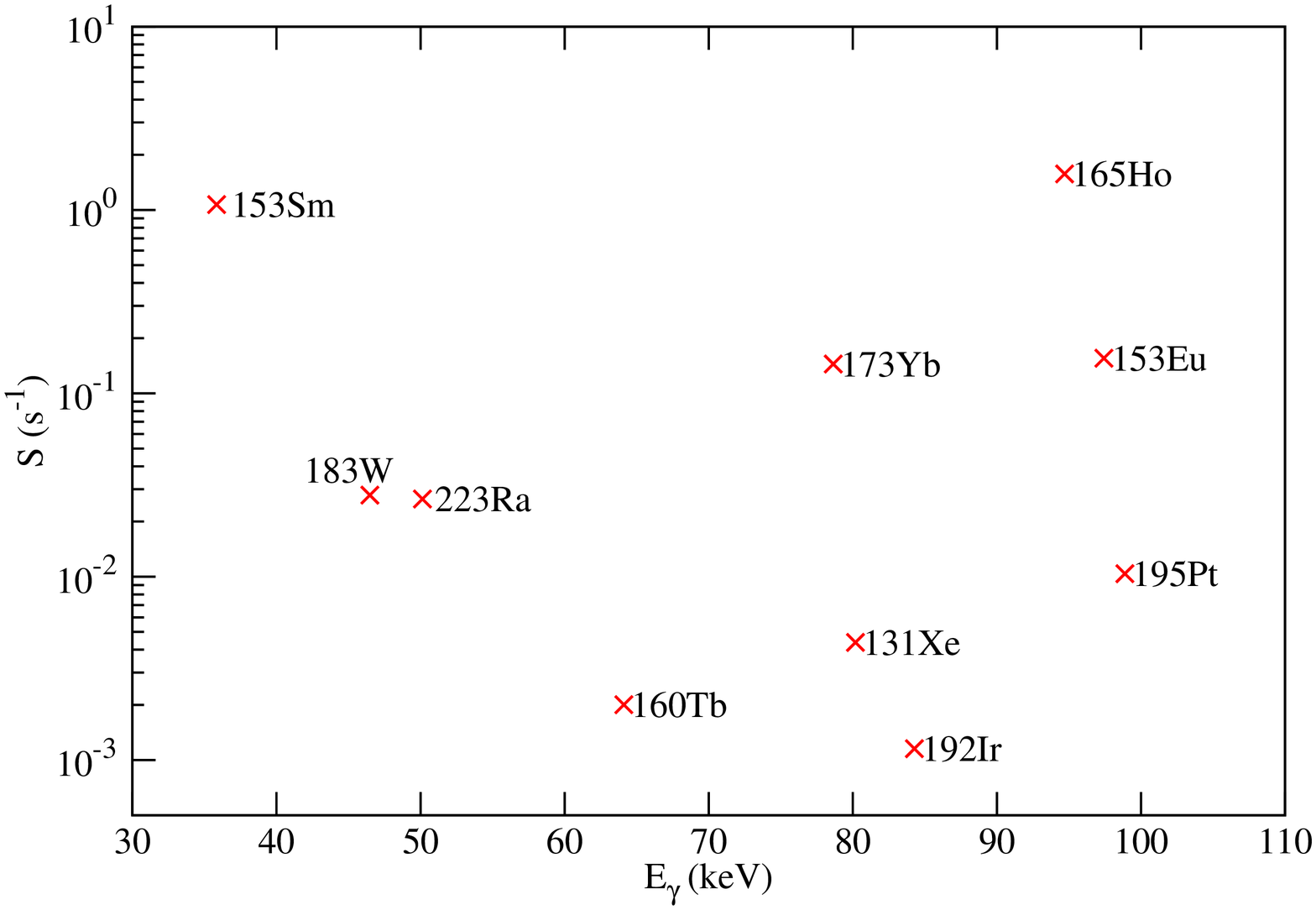}
	\end{minipage}
	%\setcaptionwidth{15cm}
	\caption{Excited state population $\rho_{ee}$ (left) and signal photon rate $S$ (right) vs. transition energy $E_{\gamma}$ obtained with XFELO parameters. The numbers are presented in Table \ref{tab:AcceleratedTarget1}.}\label{fig:NumericsRhoSAccelerated}
\end{figure}

We proceed similarly to the case of a solid state target in Sec. \ref{sec:NumericsSolidState}, but leave out IC and collective effects when determining the transition width and include the relativistic boost of all laser parameters before proceeding to the calculation of the  electric field.  Table \ref{tab:AcceleratedTarget1} shows the excited state population after one radiation pulse and the signal photon rate for European XFEL, LCLS, SACLA and XFELO laser parameters. The results for the set of parameters which yields the largest signal rates, namely for the XFELO,  are also illustrated in Fig. \ref{fig:NumericsRhoSAccelerated}.

\begin{table}[h]

\caption{ \label{tab:AcceleratedTarget1}	Excited state population $\rho_{ee}$ and signal photon rate $S$ for nuclei with transition energies above $25 \, \mathrm{keV}$ for European XFEL, LCLS, SACLA and XFELO parameters. Acceleration of the target nuclei is considered. The numbers in brackets denote the power of ten to  multiply with. See text for further explanations.}
		
\scriptsize

	\begin{tabular}{|lrc|cc|cc|cc|cc|}
 \br
 		    & $E_{\gamma}\ \, $ &  & \multicolumn{2}{|c|}{European XFEL} & \multicolumn{2}{|c|}{LCLS} & \multicolumn{2}{|c|}{SACLA} & \multicolumn{2}{|c|}{XFELO} \\ 

Nuclide &  (keV) &$\lambda $L& $\rho_{ee}$ & S ($s^{-1}$) & $\rho_{ee}$ &  S ($s^{-1}$) & $\rho_{ee}$ &  S ($s^{-1}$) & $\rho_{ee}$ &  S ($s^{-1}$) \\ 
\mr

$^{153}\mathrm{Sm}$  & 35.843 & E1 & 2.13(-7) & 2.1(-3) & $2.47(-6)$ & 7.6(-6)  & 1.11(-8) & 1.14(-8)& $4.30(-6)$ & $1.07$ \\ 
$^{183}\mathrm{W}$  & 46.484 & M1 & $5.55(-9)$ & $5.47(-5)$ & $6.43(-8)$ & $1.98(-7)$ & $1.55(-9)$ & $1.59(-9)$ & $1.12(-7)$ & $2.78(-2)$ \\ 
$^{223}\mathrm{Ra}$   & 50.128 & E1 & $5.29(-9)$ & $5.22(-5)$ & $6.13(-8)$ & $1.88(-7)$ & $2.62(-10)$ & $2.68(-10)$ & $1.07(-7)$ & $2.65(-2)$ \\
$^{160}\mathrm{Tb}$  & 64.110 & E1 & $3.99(-10)$ & $3.94(-6)$ & $4.63(-9)$ & $1.42(-8)$ & $2.30(-11)$ & $2.36(-11)$ & $8.05(-9)$ & $2.00(-3)$ \\
$^{173}\mathrm{Yb}$   & 78.647 & M1 & $2.88(-8)$ & $2.84(-4)$ & $3.34(-7)$ & $1.03(-6)$ & $6.69(-9)$ & $6.85(-9)$ & $5.81(-7)$ & $1.45(-1)$ \\
$^{131}\mathrm{Xe}$ & 80.185 & M1 & $8.73(-10)$ & $8.62(-6)$ & $1.01(-8)$ & $3.11(-8)$ & $6.50(-11)$ & $6.67(-11)$ & $1.76(-8)$ & $4.38(-3)$ \\
$^{192}\mathrm{Ir}$  & 84.275 & E1 & $2.30(-10)$ & $2.27(-6)$ & $2.67(-9)$ & $8.2(-9)$ & $1.08(-11)$ & $1.11(-11)$ & $4.64(-9)$ & $1.15(-3)$ \\ 
$^{165}\mathrm{Ho}$  & 94.700 & M1 & $3.14(-7)$ & $3.10(-3)$ & $3.64(-6)$ & $1.12(-5)$ & $3.75(-8)$ & $3.85(-8)$ & $6.33(-6)$ & $1.57$ \\ 
$^{153}\mathrm{Eu}$   & 97.429 & E1 & $3.10(-8)$ & $3.06(-4)$ & $3.59(-7)$ & $1.10(-6)$ & $1.17(-9)$ & $1.20(-9)$ & $6.24(-7)$ & $1.55(-1)$ \\ 
$^{195}\mathrm{Pt}$   & 98.882 & M1 & $2.07(-9)$ & $2.04(-5)$ & $2.40(-8)$ & $7.38(-8)$ & $4.72(-10)$ & $4.83(-10)$ & $4.18(-8)$ & $1.04(-2)$ \\
 \br
\end{tabular}

\normalsize
\end{table}

The highest excited state populations involve the $^{165}_{67} \mathrm{Ho}$ isotope. One reaches  $\rho_{ee}=3.14\times10^{-7}$ (European XFEL),  $\rho_{ee}=3.64\times10^{-6}$ (LCLS), $\rho_{ee}=3.75\times10^{-8}$ (SACLA) and $\rho_{ee}=6.33\times10^{-6}$ (XFELO). Our results are roughly three (European XFEL), four (XFELO), five (LCLS), and two (SACLA) orders of magnitude larger compared to previous estimates in Ref. \cite{Adriana_DipoleForbidden}. The rather modest results obtained for the SACLA parameters can be again traced back to the present very short pulse (10~fs) and low repetition rate of 10 Hz.\\

In general, nuclear transitions widths are larger for higher excitation energies. Furthermore, the boosting of the electric field amplitude enhances the radiation intensity in the nuclear rest frame. Therefore, we would expect that in comparison to solid state targets the disappearance of the internal conversion decay channel and of the collective effects is compensated. However, this is not the case. The low excited state populations in the case of target acceleration can be traced back to the relative uncertainty in the relativistic $\gamma$ factor, a property of the ion accelerator producing the beam that causes the bandwidth of the photons to be higher than the actual laser  bandwidth in the nuclear rest frame,  see Eq. (\ref{eq:Bandwidth}). Ideally, the relative  uncertainty  $\Delta\gamma/\gamma$ is smaller than the laser bandwidth. This is unfortunately not the case and the main advantages of the seeded XFEL and XFELO, the low bandwidth of about $10^{-7}$,  near to the Fourier width of the laser pulses, can not be exploited. 
We thus obtain overall lower excited state populations than in Sec. \ref{sec:NumericsSolidState}. Moreover, ion beams have rather small densities on the order of $\rho=10^{11} \, \mathrm{ions}/\mathrm{cm}^3$ \cite{Tahir}, which results in significantly lower signal photon rates than for solid state targets.\\

On the one hand ion beams are less problematic than solid state targets because they contain no electrons, therefore no radiation is absorbed due to the photoelectric effect as in crystals, also causing no background in the photon detector. Furthermore, they are comparatively easy to treat theoretically because they consist of bare nuclei without an electron shell. On the other hand a large accelerator is needed to produce such an ion beam of high quality and unfortunately such a facility does not exist in conjunction with an XFEL today. Furthermore, particle densities in ion beams are very low and high densities as in a solid are useful to obtain a large number of excited nuclei that we want to measure. In summary, accelerated nuclear targets in conjunction with XFEL radiation are a gedankenexperiment at the moment; nevertheless  both types of facilities are already existing, and might even be put together in the near future, for instance in the framework of MaRIE \cite{MaRIE}.

%%%%%%%%%%%%%%%%%%%%%%%%%%%%%%%%%%%%%%%%%%%%%%%%%%%%%%%%%
\section{Conclusions}
%%%%%%%%%%%%%%%%%%%%%%%%%%%%%%%%%%%%%%%%%%%%%%%%%%%%%%%%%

The interaction between nuclei and intense x-ray laser fields has been revisited to include also 
nuclear solid-state effects arising from the delocalization of the excitation over a large number of identical nuclei. 
The formation of this nuclear exciton leads to a broadening of the nuclear width which in turn increases the number of 
laser photons in resonance and the amount of excitation. The necessary conditions and limitations, mainly related to  
scattering and absorption in solid states and the laser Rayleigh length,  were investigated.  We find that by far the most promising 
candidate for laser-nucleus interaction is the well-known $^{57}\mathrm{Fe}$  M\"ossbauer isotope, for which cooperative effects can
increase the coupling of the 14.4 keV nuclear transition to the XFEL by roughly two orders of magnitude. Additionally, we have also provided an update of previous values for laser-nucleus interaction in the accelerated ion beam target setup. These values are however less promising than the results for solid-state targets.
After a long career in M\"ossbauer spectroscopy, with the 
operation of XFEL machines to produce photons of suitable energy, $^{57}\mathrm{Fe}$ might be also the first isotope to open the field of nuclear quantum optics. 
Experimentally, nuclear quantum optics has the potential to play an important role in future XFEL applications, as long-term objectives involve exciting goals such as  the preparation of nuclei in excited states, nuclear branching ratio control and isomer triggering.

%%%%%%%%%%%%%%%%%%%%%%%%%%%%%%%%%%%%%%%%%%%%%%%%%%%%%%%%%%
%\section*{Acknowledgements}
%%%%%%%%%%%%%%%%%%%%%%%%%%%%%%%%%%%%%%%%%%%%%%%%%%%%%%%%%%
% for now nobody to mention, I would say...

%%%%%%%%%%%%%%%%%%%%%%%%%%%%%%%%%%%%%%%%%%%%%%%%%%%%%%%%
%%%%%%%%%%%%%%%%%%%%%%%%%%%%%%%%%%%%%%%%%%%%%%%%%%%%%%%%
\section*{References}
%%%%%%%%%%%%%%%%%%%%%%%%%%%%%%%%%%%%%%%%%%%%%%%%%%%%5555
\bibliographystyle{iopart-num}
\bibliography{Bibliography}

\end{document}